\documentstyle[12pt]{article}
  \makeatletter
        \def\section{\@startsection {section}{1}{\z@}{3.5ex plus -1ex minus
        -.2ex}{2.3ex plus .2ex}{\large\bf}}
  \makeatother
\def\Dslash{D\!\!\!\!/\,}
\begin{document}
\setlength{\baselineskip}{22pt}
\rightline{ITP-SB-94-31}
\rightline{June 1994}
\vspace{1.3truecm}
\centerline{\large \bf Heat Kernel for Spin-3/2 Rarita-Schwinger Field}
\centerline{\large \bf in General Covariant Gauge}
\vspace{2truecm}
\renewcommand{\thefootnote}{\fnsymbol{footnote}}
\centerline{Ryusuke Endo\footnote[2]{
                                   On leave from Department of
                                   Physics, Yamagata University,
                                   Yamagata, 990 Japan.}}
\bigskip
\renewcommand{\thefootnote}{\arabic{footnote}}
\setcounter{footnote}{0}
\centerline{\it Institute for Theoretical Physics}
\centerline{\it State University of New York at Stony Brook}
\centerline{\it Stony Brook, New York 11794-3840}
\bigskip
\vspace{2truecm}
\centerline{\bf Abstract}
\bigskip
The heat kernel for the spin-3/2 Rarita-Schwinger gauge field on an
arbitrary Ricci flat space-time ($d>2$) is investigated in
a family of covariant gauges with one gauge parameter $\alpha$.
The $\alpha$-dependent term of the kernel is expressed
by the spin-1/2 heat kernel.
It is shown that the axial anomaly and the one-loop
divegence of the action are $\alpha$-independent, and that the conformal
anomaly has an $\alpha$-dependent total derivative term in $d=2m\geq6$
dimensions.

\thispagestyle{empty}
%
          \newpage
\section{Introduction}
The heat kernel and its Schwinger-De Witt expansion [1-3]
are useful tools in quantum field theory.
For example, by using them we can obtain one-loop divergent terms of
effective action \cite{DeWitt} and various kinds of anomalies
[4-6].

The heat kernels of various fields have been studied well so far
for minimal second order differential operators, that is, for the
operators whose leading term is the laplacian \cite{DeWitt,Gilkey}.
For this kind of
heat  kernel, there is an algorithm by De Witt \cite{DeWitt} of getting
expansion coefficients of coincidence limit of the kernel;
calculations can be performed covariantly at all stages. This technique
is based on the De~Witt ansatz \cite{DeWitt} which does not
apply to non-minimal operators. In gauge theories, however,
field operators are non-minimal unless special gauges are chosen.
Thus, we must deal with non-minimal operators in general.

There are several works of investigating the heat kernels for non-minimal
operators [7-11].
And there exist general
algorithms of getting the expansion coefficients covariantly
\cite{Bravinsky,GGR}. The calculations are more complicated than the
De Witt's algorithm.
Fortunately, if we restrict ourselves to the non-minimal operators
appearing in gauge theories,
there exists easier way to get the heat kernel expansion coefficients.
For the electromagnetic field $A_\mu$ in curved space-time, the present author
\cite{RE}
have given a formula such that the heat kernel in a general covariant gauge
is expressed in terms of the kernel in the Feynman gauge.
The field operator $H(\alpha)$ for $A_\mu$ in the covariant gauge
is given by
\begin{equation}
    H(\alpha) A_\mu = [ \Box \delta_\mu^\nu
                    + ({1/\alpha} - 1) D_\mu D^\nu
                    + R_\mu^{~\nu}
                      ] A_\nu ,
\end{equation}
where $\alpha$ is a numerical gauge parameter, $D_\mu$ denotes
the gravitational covariant derivative, $\Box=D_\mu D^\mu$
and $R_{\mu\nu}$ is the Ricci tensor. The operator $H(\alpha)$
is non-minimal unless $\alpha=1$ (Feynman gauge). The formula
given in ref.\cite{RE} is
\begin{equation}
  \langle x,\mu|e^{-tH(\alpha)}|x',\nu \rangle
                  = \langle x,\mu|e^{-tH(1)}|x',\nu \rangle
                  + \int_{t/\alpha}^t d\tau
                     D_\mu D^\lambda \langle
                                          x,\lambda|e^{-\tau H(1)}|x',\nu
                                     \rangle .
\label{Am}
\end{equation}
(The second term can be expressed also by the heat kernel for a scalar field
since the following identity holds \cite{RE}:
\begin{equation}
       D^\mu \langle x,\mu|e^{-tH(1)}|x',\nu \rangle
                      = - D_\nu^\prime \langle x|e^{-tH_0}|x' \rangle ,
\label{IDAm}
\end{equation}
where $H_0=\Box$  for the scalar field.)
The formula (\ref{Am}) connects the kernel for
$H(\alpha)$ to that for minimal operator $H(1)$ (and $H_0$).
Thus the expansion
coefficients can be obtained by De Witt's algorithm.
Similar formulae are given in ref.\cite{RE2} for
the antisymmetric tensor gauge fields of rank two and three.

In this paper we report a formula similar to (\ref{Am}) for the spin-3/2
Rarita-Schwinger (RS) field in arbitrary Ricci flat background curved
space-time. As applications, we discuss the gauge dependence of
one-loop divergence of effective action and axial and conformal anomalies.
The result is that the axial anomaly and the one-loop divergence are gauge
independent, while the conformal anomaly (more exactly,
$\zeta$-regularized anomalous jacobian of path integral
with respect to the local Weyl transformation) have a gauge dependent
total derivative term in $d=2m\geq6$ dimensions.

\section{Heat kernel in a general covariant gauge}
The classical lagrangian of the massless RS field $\psi_\mu$ in $d$ ($>2$)
dimensional curved space-time with metric $g_{\mu\nu}$%
               \footnote{The Euclidean signature of the metric is assumed
                         in this paper. In our convention,
                         sig$g_{\mu\nu}=(-,-,...,-)$,
                         $\gamma_\mu^\dagger = - \gamma_\mu$, and
                         the Dirac operator $\Dslash$ is hermitian.
                         }
is given by
\begin{equation}
  L_{\rm RS} = -i \sqrt{g} \bar\psi_\mu \gamma^{\mu\nu\lambda}
                             D_\nu \psi_\lambda,
\label{LRS}
\end{equation}
where $g=|\det g_{\mu\nu}|$ and $\gamma^{\mu\nu\lambda}$ is the Dirac
matrix $\gamma^\mu \gamma^\nu \gamma^\lambda$ antisymmetrized with respect
to the indices $\mu,\nu$ and $\lambda$. For the consistency we assume that the
background space-time satisfies the vacuum Einstein equation $R_{\mu\nu}=0$.
Then, the action integral is invariant under the gauge transformation
$
%
\delta \psi_\mu = D_\mu \Lambda
%
$
with arbitrary spinor $\Lambda$. To (\ref{LRS}) we add the following gauge
fixing term \cite{vanNieu}:
\begin{equation}
       L_{\rm GF} = {i \over \alpha}\sqrt{g}
                       (\bar\psi_\mu\gamma^\mu)
                       \Dslash (\gamma^\nu\psi_\nu),
\label{LGF}
\end{equation}
where $\alpha$ is a numerical gauge parameter.
Then, the total lagrangian can be written as
\begin{equation}
      L_{\rm RS} + L_{\rm GF} =
           -i \sqrt{g} \bar\phi^\mu
                  {\cal D}(\alpha^\prime)_\mu^{~\nu}
                        \phi_\nu,
\label{Ltotal}
\end{equation}
where we have made a field redefinition \cite{Takao}
\begin{equation}
   \phi_\mu = \psi_\mu - {1\over 2}\gamma_\mu \gamma^\lambda\psi_\lambda,
\label{redefinition}
\end{equation}
and the operator ${\cal D}(\alpha^\prime)$ is defined by
\begin{equation}
 {\cal D}(\alpha^\prime)_\mu^{~\nu} =
                  \Dslash \delta_\mu^\nu
                  + {2\alpha^\prime \over d}\gamma_\mu \Dslash \gamma^\nu
\label{Dalpha}
\end{equation}
with $\alpha'$ given by
\begin{equation}
    {2\alpha^\prime \over d} = {4\over (d-2)^2} \left(
                           {d-2 \over 4} - {1\over \alpha}
                           \right).
%
\end{equation}
In terms of the redefined field $\phi_\mu$, our gauge family includes
`Feynman gauge' \cite{Takao}. Namely, by choosing $\alpha=4/(d-2)$
(that is, $\alpha^\prime=0$) we get Dirac-like operator
$
         {\cal D}(0)= \Dslash,
$
the square of which is a minimal operator.
For any other value of $\alpha'$,
${{\cal D}(\alpha')}^2$ is non-minimal.

In this paper we consider the heat kernel for the $\phi_\mu$ field, that is,
the kernel for the non-minimal operator ${\cal D}(\alpha^\prime)^2$,
\begin{equation}
      K_{\mu\nu}(x,x';t) = \langle x,\mu|e^{-t{\cal D}(\alpha^\prime)^2}|
                                     x',\nu
                           \rangle
\label{Kmn}
\end{equation}
with general value of the parameter $\alpha^\prime$
($\neq \frac{d}{2(d-2)}$).
For the comparison, we put a `hat' on the kernel in the Feynman gauge:
$$
     \hat K_{\mu\nu}(x,x';t) = K_{\mu\nu}(x,x';t)|_{\alpha'=0}.
$$
In the following, we also need the spin-1/2 heat kernel $K(x,x';t)$,
\begin{equation}
      K(x,x';t) = \langle x|e^{-t {\Dslash}^2}|x'\rangle.
%
\end{equation}
We can employ the De~Witt ansatz for
$\hat K_{\mu\nu}(x,x';t)$ and $K(x,x';t)$.
For example, $K(x,x';t)$ can be expanded as \cite{DeWitt}
\begin{equation}
   K(x,x';t) = {\Delta^{1/2} \over (4\pi t)^{d/2} }
               e^{\sigma/2t}
               \sum_{n=0}^\infty b_n(x,x')t^n,
\label{1/2SDW}
\end{equation}
where $\sigma$ is the geodetic interval
and $\Delta$ is the Van Vleck determinant.
Contrary, as far as $x\neq x'$, the kernel $K_{\mu\nu}(x,x';t)$ does not
have an expansion similar to (\ref{1/2SDW}).
If $x=x'$, however, it
has an asymptotic expansion \cite{Gilkey},
\begin{equation}
  K_{\mu\nu}(x,x;t) = {1\over(4\pi t)^{d/2}}
                      \sum_{n=0}^\infty B_{n\mu\nu}(x)t^n.
\label{expKmn}
\end{equation}

The kernel $K_{\mu\nu}(x,x';t)$ satisfies the following heat
equation and boundary condition:
\begin{eqnarray}
     & &{\partial\over{\partial t}}K_{\mu\nu^\prime}(t)
          = - {\cal D}(\alpha^\prime)^2 K_{\mu\nu^\prime}(t)
\label{Kmneq} \\
    & &K_{\mu\nu^\prime}(0)
       = g^{-1/2}g_{\mu\nu} \delta(x-x'),
\label{Kmn0}
\end{eqnarray}
where we have used abbreviation such as
$K_{\mu\nu^\prime}(t)=K_{\mu\nu}(x,x';t)$; primed [unprimed] indices
denote vector indices at the point $x'$ [$x$].
Similar equations characterize 
$K(t)$.

Assuming now $\hat K_{\mu\nu^\prime}(t)$ satisfies (\ref{Kmneq}) and
(\ref{Kmn0}) with $\alpha'=0$,
we can write the solution to (\ref{Kmneq}) and
(\ref{Kmn0}) with general value of $\alpha'$ as
\begin{equation}
   K_{\mu\nu^\prime}(t) = \hat K_{\mu\nu^\prime}(t)
                       + \sum_{i=+,-} {1\over N_i} \int_{{l_i}^2t}^t d\tau
                 \left\{
                         D_\mu D^\lambda
                       + a_i (\gamma_\mu \Dslash D^\lambda
                               + D_\mu \Dslash \gamma^\lambda)
                       + {a_i}^2 \gamma_\mu \Dslash ^2 \gamma^\lambda
                 \right\}
                         \hat K_{\lambda\nu^\prime}(\tau)
\label{KhatK}
\end{equation}
where $N_{\pm}$, $a_{\pm}$ and $l_{\pm}$ are functions of $\alpha^\prime$
defined by
\begin{eqnarray}
           & & N_{\pm} = d {a_{\pm}}^2 + 2a_{\pm} + 1 ,\nonumber \\
           & &2a_{\pm} = l_{\pm} - 1,    \nonumber \\
           & &l_{\pm} = \alpha^\prime \pm f(\alpha'),    \nonumber \\
           & &f(\alpha') = \sqrt{ (1 - \alpha^\prime)^2
                           + 4 \alpha' /d
                             }.
\label{f(a)}
\end{eqnarray}
This is the spin-3/2 counterpart of (\ref{Am}). In contrast to the
electromagnetic case, there appear two $\tau$-integrations. This is
attributable to the fact that $\phi_\mu$ includes
two spinor components $\gamma^\mu\phi_\mu$ and $D^\mu\phi_\mu$;
while $A_\mu$ has only one scalar component $D^\mu A_\mu$.
It is straightforward to see that the right hand side of (\ref{KhatK})
satisfies (\ref{Kmneq}) and (\ref{Kmn0}) in the Ricci flat space-time.

The second term of (\ref{KhatK}) can be expressed also by the spin-1/2
heat kernel $K(t)$. In the Ricci flat space-time, we have the following
identities analogous to (\ref{IDAm}):
\begin{eqnarray}
       \gamma^\mu \hat K_{\mu\nu^\prime}(t) &=& K(t) \gamma_{\nu^\prime},
                                                        \nonumber \\
       D^\mu \hat K_{\mu\nu^\prime}(t) &=& - D_\nu^\prime K(t).
\label{KmnK}
\end{eqnarray}
Thus, (\ref{KhatK}) can be written as
\begin{eqnarray}
   K_{\mu\nu^\prime}(t) &=&
           \hat K_{\mu\nu^\prime}(t)
           - \sum_{i=+,-} {1\over N_i} \int_{{l_i}^2t}^t d\tau
      \Bigl\{
             D_\mu D_\nu^\prime K(\tau)
                                       \nonumber \\
       & &
             + a_i \Bigl(
                         \gamma_\mu \Dslash D_\nu^\prime K(\tau)
                        - D_\mu \Dslash K(\tau)\gamma_{\nu^\prime}
                   \Bigr)
             - {a_i}^2 \gamma_\mu {\Dslash}^2 K(\tau)\gamma_{\nu^\prime}
      \Bigr\}
\label{KmnK1/2}
\end{eqnarray}

The expression (\ref{KhatK}) or (\ref{KmnK1/2})
enables us to use De~Witt's technique to get the asymptotic
expansion coefficients $B_{n\mu\nu}(x)$ in (\ref{expKmn}).
{}From (\ref{KmnK1/2}) and (\ref{1/2SDW}) it follows that
\begin{eqnarray}
  B_{n\mu\nu}(x) &=& \hat B_{n\mu\nu}(x)
                    - {p_n \over 2} g_{\mu\nu} [b_n]
                    - {q_n \over 2} \left(
                                     \gamma_\mu\gamma_\nu [b_n]
                                     + [b_n] \gamma_\mu \gamma_\nu
                                    \right)
 \nonumber \\
             & &    + q_n \left(
                        \gamma_\mu\gamma^\lambda[b_{n-1;\lambda\nu'}]
                        + [b_{n-1;\mu\lambda'}]\gamma^\lambda \gamma_\nu
                          \right)
                    + r_n \gamma_\mu [b_n] \gamma_\nu,
\label{Bn}
\end{eqnarray}
where $\hat B_{n\mu\nu}(x)= B_{n\mu\nu}(x)|_{\alpha'=0}$ is
the expansion coefficient of $\hat K_{\mu\nu}(x,x;t)$, and
\begin{eqnarray}
    p_n &=& { k_{+n}\over N_+}
                 + { k_{-n}\over N_-}, \nonumber \\
    q_n &=& { a_+ k_{+ n}\over N_+}
                  + { a_- k_{- n}\over N_-}, \nonumber \\
    r_n &=& \left(
                  n-{d\over2}
            \right)
                     \left(
                             { {a_+}^2 k_{+ n}\over N_+}
                              + { {a_-}^2 k_{- n}\over N_-}
                     \right)
\label{pqrn}
\end{eqnarray}
with
\begin{equation}
   k_{\pm n} = \left\{
                   \begin{array}{cl}
                        {\displaystyle
                           { {({l_\pm}^2)}^{n-d/2 } -1
                                    \over n-d/2 }
                         }
                            , & \quad (n \neq d/2 ) \\
                                       &             \\%
                        \ln {l_\pm}^2. & \quad (n = d/2 )
                   \end{array}
               \right.
\end{equation}
In (\ref{Bn}) we have used Synge's bracket symbol \cite{Synge}
to denote the coincidence limit such as
$$
          [b_{n;\mu\nu'
                        }]
             = \lim_{x'\rightarrow x} D'_{\nu}D_\mu
                                      b_n(x,x').
$$
Expansion coefficients $[b_n]$ and $\hat B_{n\mu\nu}$ for lower $n$
are given in refs.\cite{DeWitt,Gilkey}:
\begin{eqnarray}
         [b_0] &=& {\bf 1}, \quad \quad \quad [b_1] = 0, \nonumber \\
         ~[b_2] &=& {1\over 180}R_{\alpha\beta\gamma\delta}
                               R^{\alpha\beta\gamma\delta}{\bf 1}
                  + {1\over 12}{\bf R}_{\alpha\beta}
                               {\bf R}^{\alpha\beta};
\label{[bn]} \\
   & & \nonumber \\
     \hat B_{0\mu\nu} &=& g_{\mu\nu}{\bf 1},\quad \quad
          \hat B_{1\mu\nu} = -2{\bf R}_{\mu\nu}, \nonumber \\
     \hat B_{2\mu\nu} &=&
                 {1\over 3}\Box{\bf R}_{\mu\nu}
               + {1\over180}g_{\mu\nu} R_{\alpha\beta\gamma\delta}
                                       R^{\alpha\beta\gamma\delta}{\bf 1}
               - {1\over12}  R_{\mu\alpha\beta\gamma}
                                       R_\nu^{~\alpha\beta\gamma}{\bf 1}
   \nonumber \\
           &+& {1\over6}R_{\mu\nu\alpha\beta}{\bf R}^{\alpha\beta}
               + {1\over12}g_{\mu\nu}{\bf R}_{\alpha\beta}
                                     {\bf R}^{\alpha\beta}
               - 2 {\bf R}_{\mu\alpha}{\bf R}_\nu^{~\alpha},
\label{Bhatn}
\end{eqnarray}
where we have used $R_{\mu\nu}=0$, and ${\bf R}_{\mu\nu}$ is defined by
\begin{equation}
           {\bf R}_{\mu\nu} = {1\over 8}[\gamma^\alpha, \gamma^\beta]
                                         R_{\alpha\beta\mu\nu}.
\end{equation}
The coincidence limit of
$[b_{n;\mu\nu'}]$
can be calculated by using De~Witt's technique \cite{DeWitt}:
\begin{eqnarray}
      [b_{0;\mu\nu'}] &=& {1\over 2} {\bf R}_{\mu\nu}, \nonumber \\
     ~[b_{1;\mu\nu'}] &=& {1\over 90} R_{\mu\alpha\beta\gamma}
                                      R_\nu^{~\alpha\beta\gamma}{\bf 1}
                 + {1\over 12} (
                      {\bf R}_{\mu\alpha} {\bf R}_\nu^{~\alpha}
                      + {\bf R}_{\nu\alpha} {\bf R}_\mu^{~\alpha}
                               ).
\label{bmn'}
\end{eqnarray}
To get these quantities it is convenient to employ generalized
Synge's theorem \cite{Christensen,Synge}: For any two point tensor-spinor
$T_{\alpha_1...\alpha_n\beta'_1...\beta'_m}$,
\begin{equation}
          [T_{\alpha_1...\alpha_n\beta'_1...\beta'_m}]_{;\mu}
        =  [T_{\alpha_1...\alpha_n\beta'_1...\beta'_m;\mu}]
        +  [T_{\alpha_1...\alpha_n\beta'_1...\beta'_m;\mu'}].
\label{Theorem}
\end{equation}

Substituting (\ref{[bn]}), (\ref{Bhatn}) and (\ref{bmn'}) into (\ref{Bn}),
we get
\begin{eqnarray}
   B_{0\mu\nu} &=& \left(
                         1-{p_0\over 2}
                   \right)
                          g_{\mu\nu}{\bf 1}
                   + (r_0 - q_0)\gamma_\mu \gamma_\nu,
    \nonumber \\
   B_{1\mu\nu} &=& \left(
                          -2 + {p_1\over 2}
                   \right)
                          {\bf R}_{\mu\nu}.
\label{B01}
\end{eqnarray}
The $\alpha$-dependence appears through the coefficients $p_n$, $q_n$
and $r_n$.
For example, in four dimensions,
\begin{eqnarray}
   &  &p_0 = -{1\over 2}(\alpha-2)^2
                         (\alpha^2 - 3\alpha +5),
            \nonumber \\
   &  &r_0 - q_0 =
               {1\over 8} (\alpha -2)
                          (2\alpha^2 - 5\alpha + 6),
          \nonumber \\
   &  &p_1  =  - (\alpha - 2)^2 .
\label{pqr01}
\end{eqnarray}
It is also easy to get the expression for $B_{2\mu\nu}$.
Especially in four dimensions, after using some identities
valid in $d=4$, we find
\begin{equation}
     B_{2\mu\nu} = \hat B_{2\mu\nu}. \quad \quad (d=4)
\label{B2mn}
\end{equation}
Namely,  $B_{2\mu\nu}$ is gauge independent in four dimensions.
In higher dimensions, however, it has $\alpha$-dependence:
\begin{eqnarray}
   B_{2\mu}^{~~~\mu}
   &=&
   {1\over 180} \left\{ (d-15)
                       - (d-4) \left(
                                    {p_2\over 2} + q_2
                             \right)
                        + (d - 15)r_2
                \right\}
                        R_{\alpha\beta\gamma\delta}
                        R^{\alpha\beta\gamma\delta}
     \nonumber \\
   &+&
   {1\over 12}  \left\{ (d-24)
                       - (d-4) \left(
                                    {p_2\over 2} + q_2
                               \right)
                       + (d - 8) r_2
                \right\}
                        {\bf R}_{\alpha\beta}
                        {\bf R}^{\alpha\beta}.
\label{B2mm}
\end{eqnarray}

\section{Axial and conformal anomalies, and one-loop divergence}
As applications of the formula (\ref{KmnK1/2}),
we consider here
the axial anomaly ${\cal A}_{\rm axial}(x)$,
the conformal anomaly ${\cal A}_{\rm conf}(x)$ and the one-loop
logarithmic divergent term $S_{\rm div}$ of the action in
$d=2m$ dimensional space-time. Following the standard procedure
(for example, see refs.\cite{DeWitt,Kimura,Dowker,Fujikawa}),
these quantities are given by the coefficient $B_{m\mu\nu}(x)$:
\begin{eqnarray}
  {\cal A}_{\rm axial}(x) &\propto&
                                {\rm tr}
                                 \gamma_5 B_{m\mu}^{~~~\mu}(x),
       \label{axial} \\
  {\cal A}_{\rm conf}(x) &\propto&
                                {\rm tr} B_{m\mu}^{~~~\mu}(x),
       \label{conf} \\
  S_{\rm div} & \propto &
                                    \int {\rm tr} B_{m\mu}^{~~~\mu}(x)
                                    \sqrt{g} d^{2m}x ,
        \label{div}
\end{eqnarray}
to which the contributions from the Faddeev-Popov ghosts \cite{NK}
are assumed to be added.
By the `conformal anomaly' we mean the anomalous term coming from
the path integral jacobian with respect to the local Weyl
transformation \cite{Fujikawa} of the $\phi_\mu$-field.
To regularize the anomalous jacobian, we have
used the $\zeta$-function regularization \cite{Dowker}.

As seen from (\ref{B2mn}), there are no gauge dependent term in these
quantities in four dimensions.  To get the higher
dimensional results, let us consider the quantity
$[K_\mu^{~\mu'}(t)]=K_\mu^{~\mu}(x,x;t)$.
By successive use of Synge's theorem (\ref{Theorem}) to
$K_{;\mu\nu'}(\tau)$ in (\ref{KmnK1/2}), we can write
\begin{eqnarray}
  [K_\mu^{~\mu'}(t)] &=& [\hat K_\mu^{~\mu'}(t)]
                + \sum_{i=+,-}{1\over 2N_i}
                  \int_t^{{l_i}^2t} d\tau \Box [K(\tau)]
  \nonumber \\
        &+& \sum_{i=+,-} {1\over N_i} \Bigl\{
                (1 \pm 2a_i)([K({l_i}^2t)] - [K(t)])
                + {a_i}^2 \gamma_\mu ([K({l_i}^2t)] - [K(t)])\gamma^\mu
                                     \Bigr\}
  \nonumber \\
        &+& \sum_{i=+,-}{1\over N_i} \int_t^{{l_i}^2t} d\tau a_i
                 \left\{
                        \Bigl[
                               \gamma^\mu, [D_\mu'\Dslash K(\tau)]
                        \Bigr]_\mp
                       - \Bigl[
                               [D_\mu \Dslash K(\tau)],\gamma^\mu
                         \Bigr]_\mp
                 \right\},
\label{[Kmm]}
\end{eqnarray}
where $[X,Y]_\mp=XY\mp YX$. In deriving (\ref{[Kmm]}),
we have also used the identity
\begin{equation}
        \Dslash K(t) = - K(t) \stackrel{\leftarrow}{\Dslash^\prime}
\end{equation}
together with the heat equation for $K(t)$.

Now we can consider the quantities ${\rm tr}B_{m\mu}^{~~~\mu}$
and ${\rm tr}\gamma_5 B_{m\mu}^{~~~\mu}$.
Since $B_{m\mu}^{~~~\mu}(x)$
is the $t$-independent term in the asymptotic expansion of
$K_\mu^{~\mu}(x,x;t)$, the third term of (\ref{[Kmm]})
does not contribute to $B_{m\mu}^{~~~\mu}$. The fourth term does not
contribute either after the trace is taken, because
$
       {\rm tr}(\gamma^\mu X - X \gamma^\mu)=
       {\rm tr} \gamma_5 (\gamma^\mu X + X \gamma^\mu)=0.
$
Thus the gauge dependent term comes only from the second term:
\begin{equation}
        {\rm tr} \Gamma B_{m\mu}^{~~\mu}
               = {\rm tr} \Gamma \hat B_{m\mu}^{~~\mu}
              + {p_m\over 2} \Box {\rm tr} \Gamma [b_{m-1}],
               \quad \quad (\Gamma = {\bf 1}, \gamma_5)
\label{trBmm}
\end{equation}
where $\alpha'$-dependence of $p_m$ is given by
\begin{equation}
p_m = {2m\over 2m-1} \ln \left|
                               {2(m-1)\alpha'\over m} -1
                         \right|
    - {2(m\alpha'-m+1)\over (2m-1)f(\alpha')}
                      \ln \left|
                                \alpha'+f(\alpha')
                                \over
                                \alpha'-f(\alpha')
                          \right|.
\end{equation}
with $f(\alpha')$ defined by (\ref{f(a)}).

{}From (\ref{trBmm}) we can conclude that the axial anomaly (\ref{axial}) is
$\alpha$-independent. In fact, ${\rm tr}\gamma_5 [b_{m-1}]=0$ since
$[b_{m-1}]$ has at most $2(m-1)=d-2$ factors of $\gamma$-matrices
\cite{Takao}.
This result agrees with the previous literature \cite{Takao,Gordon}.

The $\alpha$-independence of the one-loop divergence (\ref{div})
is also obvious, since the second term of (\ref{trBmm}) is total derivative.
On the contrary, the gauge dependent term does not vanish for the
conformal anomaly (\ref{conf});
tr$[b_{m-1}]$ does not vanish in general ($d\geq6$).
For example, in $d=6$ dimensions, ${\cal A}_{\rm conf}$ has a gauge dependent
term proportional to
\begin{equation}
      {p_3 \over 2} \Box {\rm tr}[b_2]
             = - {7p_3 \over 360} \Box
                     (R_{\alpha\beta\gamma\delta}
                      R^{\alpha\beta\gamma\delta}).
\label{6d}
\end{equation}

\section{Discussion}
There are three possibilities to explain the gauge dependence of the
conformal anomaly:
\begin{enumerate}
\item  The massless spin-3/2 system is not conformal invariant
       even on the classical level.
       In such a system, the trace of the energy-momentum tensor is
       not equal to the conformal anomaly. Instead,
       \begin{equation}
          \langle T_\mu^{~\mu} \rangle = {\cal A}_{\rm conf}
                  + D_\mu \langle J^\mu \rangle,
       \label{Tmm}
       \end{equation}
       with a current $J^\mu(x)$ \cite{Fujikawa,Grisaru}. This current
       may show gauge dependence so that
       the `physical' $\langle T_\mu^{~\mu} \rangle$ may
       have no gauge dependence \cite{Grisaru}.
\item  Even classically, the energy-momentum tensor of spin-3/2
       system depends on gauge \cite{Das}.
       Thus, the left hand side of (\ref{Tmm}) may also
       show gauge dependence.
\item  When we introduce the gauge fixing term $L_{\rm GF}$ (\ref{LGF}),
       we should also consider the functional determinant
       $\det \alpha \delta(x-x'){\bf 1}$
       in the path integral \cite{vanNieu2}. (This factor may be
       included in the ghost determinant \cite{vanNieu2}.)
       As in the case of vector and antisymmetric
       tensor gauge fields \cite{vanNieu2}, this factor may yields a
       gauge dependent contribution to $\langle T_{\mu}^{~\mu} \rangle$.
\end{enumerate}
These three possible $\alpha$-dependences should balance
the dependence from ${\cal A}_{\rm conf}$ in (\ref{Tmm});
further study is needed to check it.
We emphasize here that, besides the possibilities mentioned above,
the gauge dependent term (\ref{6d}) of the anomaly in six dimensions
can be removed by a finite counterterm. In fact, we can see
\begin{equation}
                  \sqrt{g} \Box (
                              R_{\alpha\beta\gamma\delta}
                              R^{\alpha\beta\gamma\delta}
                                           )
                   = -{2\over3}
               g_{\mu\nu}{\delta \over \delta g_{\mu\nu}}
                \int d^6x \sqrt{g} R_{\mu\nu}^{~~~\alpha\beta}
                      R_{\alpha\beta}^{~~~\rho\sigma}
                      R_{\rho\sigma}^{~~~\mu\nu}
\end{equation}
in Ricci flat space-time.  Moreover, if we allow more
artificial counterterms, $\Box {\rm tr}[b_{m-1}]$ in $2m$ dimensions
can be removed, since it may be written as
\begin{equation}
                \sqrt{g} \Box {\rm tr}[b_{m-1}]
                = - \left.
                {1 \over m-1}
                g_{\mu\nu}{\delta \over \delta g_{\mu\nu}}
                     \int d^{2m}x \sqrt{g}
                      {\rm tr}[b_{m-1}]
                      R
               \right|_{R_{\mu\nu}=0}.
\end{equation}

\section{Acknowledgements}
I would like to thank the members of the Institute for Theoretical
Physics at Stony Brook for their hospitality.
I am also grateful to Peter van Nieuwenhuizen for discussion.
This work was supported in part by NSF Grant \#PHY9309888.
\setlength{\baselineskip}{18pt}

%
%
\end{document}